\documentstyle[12pt]{article}
\def \nubar {\overline {\nu}}
\def \osctau  {{$\nu_\mu \leftrightarrow \nu_\tau$}}
\def \oscster {{$\nu_\mu \leftrightarrow \nu_s$}}
\def \oscel   {{$\nu_\mu \leftrightarrow  \nu_e$}}
\def \blankline{\vspace{0.4 cm}}
\def\aprle{\buildrel < \over {_{\sim}}}
\def\aprge{\buildrel > \over {_{\sim}}}

\begin{document}

\section*{\centerline
{Matter Effects in Atmospheric Neutrino Oscillations}}

\blankline
\centerline {E. Akhmedov\footnote{On leave from Kurchatov Institute of
Atomic Energy, Moscow 123182, Russia}, P. Lipari and M. Lusignoli}

\blankline
\centerline {\it Dipartimento di Fisica, Universit\`a di Roma ``la Sapienza",}
\centerline {\it and I.N.F.N., Sezione di Roma,}
\centerline {\it Piazzale A. Moro 2,   I-00185 Roma, Italy}

\blankline
\begin{abstract}
The Kamiokande II and IMB data on contained events induced by atmospheric
neutrinos exhibit too low a ratio of muons to electrons, which has been
interpreted as a possible indication of neutrino oscillations. At the same
time, the recent data on upward--going muons in underground detectors have
shown no evidence for neutrino oscillations, strongly limiting the allowed
region of oscillation parameter space. In this paper we confront
different types of neutrino oscillation hypotheses with the experimental
results.
 The matter effects in $\nu_\mu
\leftrightarrow \nu_e$ and in $\nu_\mu
\leftrightarrow \nu_{sterile}$ oscillations are discussed and
shown to affect significantly the upward--going muons.
\end{abstract}

\vspace{\baselineskip}

Several times signals of neutrino oscillations have been claimed in
the past. The evidence from accelerator and reactor
experiments has meanwhile evaporated \cite{ACCOSC}, leaving
the observed solar neutrino deficit as the only possible indication.

Recently, however, another hint of neutrino oscillations has
been obtained.
The Kamiokande II \cite{KAM92a} and IMB \cite{IMB92a}
collaborations observed an anomaly in the contained neutrino
induced events. Both groups measure the ratio of muons to electrons
and find it smaller than what is predicted by
several independent   theoretical calculations \cite{contained-theory}.
The events observed in these experiments have charged lepton energies
in the range from a few hundred MeV to 1.2 GeV. A possible interpretation
of the effect
has been given in terms of $\nu_\mu \leftrightarrow \nu_x$
oscillations, where $\nu_x$ can be  $\nu_e$, $\nu_\tau$ or a
sterile neutrino $(\nu_s)$, with
typical values  for the oscillation parameters  $\Delta m^2 \sim 10^{-3}
\div 10^{-2}~{\rm eV^2}$ and $\sin^2 2\theta  \ge 0.5$
(a summary is contained in ref. \cite{KAM92c}).
For this range of parameters, the flux of upward--going
muons could  also be observably reduced. However, the IMB and Baksan
experiments  did not observe any such reduction \cite{IMB92b,Baksan91}, thus
setting rather stringent limits to the allowed region in the  ($\Delta m^2$,
$\sin^2 2\theta$) plane.

Neutrino oscillations
(in vacuum \cite{Pontecorvo} or in solar matter
\cite{MSW}) are also   the most popular explanation for the
solar neutrino deficit.
In this case the $\nu_e$ must  oscillate into a  neutrino
of different flavour. However, both problems cannot be solved
through the same $\nu_e \leftrightarrow \nu_\mu$ oscillations, since the
ranges of $\Delta m^2$ required  in  the two cases do not overlap.
Thus, if the solar neutrino problem is solved through
$\nu_e \leftrightarrow \nu_\mu$ oscillations, then
the atmospheric $\nu_\mu$'s must oscillate into something else.
The simplest and often  made hypothesis is
that the  contained events anomaly is explained through the
$\nu_\mu \leftrightarrow \nu_\tau$  oscillations.
In this  case, since both neutrino flavours have identical interactions
with matter,  the  oscillations will not  be affected by the presence of
the matter in  the earth. If, however, the initial $\nu_\mu$ oscillates to
$\nu_e$  or to a  sterile neutrino, $\nu_s$, the matter effects can be
relevant.

In this letter we shall discuss in detail the various scenarios for
atmospheric neutrino oscillations and show that indeed the matter effects
are important,  in that the allowed parameter region
for both  $\nu_\mu \leftrightarrow \nu_e$ and
$\nu_\mu \leftrightarrow \nu_s$ oscillations  is appreciably
different as    compared to the $\nu_\mu \leftrightarrow \nu_\tau$ case.

\vspace{0.4 cm}
The evolution equation \cite{MSW} of a system of two neutrino
species ($\nu_i, \nu_j$) in matter is
\begin{equation}
i\frac{d}{dt}\left(\begin{array}{l}
   \nu_{i}\\
   \nu_{j}
\end{array}\right )
{}~=~   \frac{\Delta m^2}{4E} \left ( \begin{array}{cc}
  2 v_{i}(t)-\cos 2\theta & \sin 2\theta\\
   \sin 2\theta & 2 v_{j}(t)+\cos 2\theta
\end{array}\right)
\left(\begin{array}{l}
   \nu_{i}\\
   \nu_{j}
\end{array}\right )
\label{evolution}
\end{equation}
Here $i,j=e,\mu,\tau$ or $s$ where $\nu_s$ is a sterile neutrino, $i\not=
j$, $\theta$ is the $\nu_{i}$---$\nu_j$ mixing angle in vacuum
($0 \le \theta \le \pi/4$), $\Delta m^2=
m_2^2-m_1^2$, where we assume $\nu_i$ to be a predominant component of the mass
eigenstate $\nu_1$, and
\begin{equation}
v_{e}(t)  =   {2\sqrt{2}EG_{F} \over \Delta m^2} ~
\left [ N_{e}(t)-{N_n(t)\over 2} \right ],
\label{v_e}
\end{equation}
\begin{equation}
v_{\mu}(t)  =  v_{\tau} (t) =  {2\sqrt{2}EG_{F} \over \Delta m^2} ~
\left [-{N_n(t)\over 2} \right ],
\label{v_mu}
\end{equation}
\begin{equation}
v_{s}(t)  =   0
\label{v_s}
\end{equation}
where $N_{e}$ and $N_{n}$ are the electron and neutron number densities
and $G_{F}$ is the Fermi constant.
The evolution of the antineutrino system  is described by
the same equation
with  $v_{i,j}(t)$ substituted by $-v_{i,j}(t)$.

In matter of constant density the probability of $\nu_i \leftrightarrow
\nu_j$ oscillations has the same form as in vacuum with the amplitude
$\sin^2 2\theta$ and oscillation length $\ell=4\pi E/\Delta m^2$
substituted by
\begin{equation}
\sin^2 2\theta_m =\frac{\sin^2 2\theta}{[(v_i-v_j)-\cos 2\theta]^2+
\sin^2 2\theta},
\label {mix-mat}
\end{equation}
\begin{equation}
\ell_m = {\ell \over
\{ [(v_i-v_j)-\cos 2\theta]^2+\sin^2 2\theta \}^{1/2} }
\end{equation}
It can be seen from eq.(\ref{mix-mat})
 that matter can either enhance or suppress
oscillations depending on the relative  signs and magnitudes of $(v_i-v_j)$,
and $\cos 2\theta$.
Note that
$(v_i - v_j)$ changes its sign   with
$\Delta m^2$.
If the condition
\begin{equation}
v_i-v_j=\cos 2\theta
\end{equation}
is satisfied, the amplitude of oscillations in matter $\sin^2 2\theta_m$
becomes equal to unity, and the oscillations are resonantly enhanced
(the MSW effect).
The probability of the $\nu_i \rightarrow \nu_j$ transition will be large
provided matter density varies slowly enough along the neutrino trajectory,
i.e. the adiabaticity condition is fulfilled.
In matter with uniform (or nearly uniform) density the transition
probability will crucially depend on the thickness of the matter slab.
In general one needs to integrate numerically  the evolution equation
(\ref{evolution}) along the neutrino path.

\vspace {0.5 cm}
The effect of matter on  $\sin^2 2 \theta_m$
in the case of $\nu_\mu \leftrightarrow \nu_e$  oscillations
is illustrated in fig. 1, where
it is plotted as a function
of $\Delta m^2/E$,  assuming
a density $\rho = 5~{\rm g\,cm^{-3}}$, and an electron fraction
$N_e/(N_p + N_n) = 1/2$
(this corresponds to approximately average values in the earth).
We can distinguish  three regions:
if $\Delta m^2/E \aprge  10^{-2}~{\rm eV^2/GeV}$, the matter effects are
negligible;
if $\Delta m^2/E \aprle 10^{-4}~{\rm eV^2/GeV}$,
the matter effect supresses strongly the  oscillation probabilities,
and mixing in matter almost vanishes; in the transition region
for $\Delta m^2 > 0$
($\Delta m^2 < 0$)
the neutrino (antineutrino) mixing parameter passes through a resonance
$\sin^2(2 \theta_m) = 1$  of width proportional to  $\tan 2 \theta$,
while the
antineutrino (neutrino) effective mixing parameter goes to zero monotonically
as the energy increases.
In the case of maximal mixing in vacuum
the oscillation probabilities
of  neutrinos and antineutrinos are equal
and the sign of $\Delta m^2$ is irrelevant.

The  case of $\nu_\mu \leftrightarrow \nu_s$ oscillations
can be calculated as  before
with the substitution
$N_e(t) \to -N_n(t)/2$. The change of sign has the consequence that
now for $\Delta m^2 > 0$
($\Delta m^2 < 0$)  it is the antineutrino (neutrino) that will pass through
the resonance.
Since in the earth
$N_n \simeq  N_p = N_e$,
fig. 1 can also be interpreted as  describing the
dependence on $\Delta m^2/E$ of  the mixing parameter $\sin^2(2\theta_m)$ in
the case of $\nu_\mu \leftrightarrow \nu_s$  oscillations  in matter  with
a density of 10 ${\rm g\,cm^{-3}}$, interchanging the neutrino and
antineutrino curves.

\vspace{0.5 cm}
Atmospheric neutrinos are produced in the hadronic shower induced by primary
cosmic rays in the earth's atmosphere. These neutrinos can be observed
directly in large mass underground detectors by means of their charged-current
interactions. ``Contained events'' are those where the neutrino--nucleus
interaction vertex is located inside the detector
and all final state particles do not get out from it. The average energy of
the neutrinos that give rise to these events is a few hundred
MeV.

Muon neutrinos can also be detected indirectly observing the muons that they
have produced in the material surrounding the detector. To reduce the
background from atmospheric muons, only upward--going neutrino-induced muons
are usually considered. A rough estimate of the energy spectrum of the
upward--going muons has been obtained dividing them in two categories,
passing (those of highest energy) and stopping muons.

\vspace{0.5 cm}
The flux of upward--going muons can be calculated as
\begin{equation}
\phi_{\mu^\mp} (E_\mu, \cos \psi) = \int_{E_\mu}^{\infty} dE_\nu
{}~\phi_{\nu_\mu(\nubar_\mu)} (E_\nu, \cos \psi)
{}~{dn_{\nu_\mu(\nubar_\mu) \to \mu^\mp} \over dE_{\mu} }
(E_\mu;E_\nu)
\label{upw-flux}
\end{equation}
where $\phi_{\nu_\mu(\nubar_\mu)}(E_\nu, \cos\psi)$ is the $\nu_\mu$
($\nubar_\mu$) spectrum at  nadir angle
$\psi$, and  $dn_{\nu_\mu(\nubar_\mu) \to \mu^\mp}/dE_\mu$
is the number of $\mu^\mp$  with energy  between
$E_\mu$ and $E_\mu + dE_\mu$ produced by a
neutrino  (antineutrino) of energy $E_\nu$:
\begin{equation}
{dn_{\nu_\mu(\nubar_\mu) \to \mu^\mp} \over dE_{\mu} } (E_\mu;E_\nu) =
N_A \int_0^\infty dX~\int_{E_\mu}^{E_\nu} dE_0
{d \sigma_{\nu_\mu(\nubar_\mu)} \over dE_0} (E_0, E_\nu)
 \delta  [E_\mu - R^{-1}(R(E_0) - X) ]
\end{equation}
In this equation,
the first integral is over the  position of  neutrino absorption
($X$ is in ${\rm g\,cm^{-2}}$, $X=0$ corresponds to the detector);
the second  integral is
over the  muon energy at the production point.
The delta function expresses the fact that the muon created  with
initial energy $E_0$ at the point $X$  must  reach
the detector with energy $E_\mu$.
In the argument of the delta function,
$R(E)$ is the range in  rock of a muon  of energy $E$, and
$R^{-1}(X)$ is the inverse  function, giving the initial energy of a
muon with range $X$.
We are  assuming that fluctuations in the energy loss
are negligible, and that the energy--range relation is well defined.
This is a good approximation in this range of  muon energy
(see ref. \cite{LS91} for a discussion).
The neutrino cross section,
expressed in terms of the
the usual  kinematical variables
$ y = 1 - E_\mu/E_\nu$, and $x = Q^2/(2 m_N E_\nu y)$,
can  be calculated  integrating over $x$ the following expression:
\begin{equation}
{d^2 \sigma_{\nu} \over dx dy } (x,y,E_\nu) = {2G_F^2\;m_N\;E_\nu  \over \pi}
{}~\left [ {M_W^2 \over M_W^2 + Q^2} \right ]^2
{}~ x [q_{d} (x, Q^2) + (1-y)^2~\overline {q}_{u} (x,Q^2)] \label{nuxsec}
\end{equation}
In eq. (\ref{nuxsec})
$G_F$ is the Fermi constant, $M_W$ is the $W$ boson mass,  $m_N$ is the
nucleon mass, and  $q_{d,u}$  ($\overline {q}_{d,u}$ ) are the distributions
of quarks (antiquarks) of down--like and up--like type in the target
nucleon.
The antineutrino cross section  is obtained  interchanging
$q \leftrightarrow \overline{q}$.

Eq. (\ref{upw-flux}) assumes that the neutrino flux is unmodifed
in the passage through the earth.
To take into account the  effects of oscillations we have to  make
the substitution
\begin{equation}
\phi_{\nu_\mu} \to [1 - P_{\nu_\mu \to \nu_x}]~ \phi_{\nu_\mu}
\end{equation}
for $\nu_x = \nu_\tau$ or $\nu_s$, or
\begin{equation}
\phi_{\nu_\mu} \to [1 - P_{\nu_\mu \to \nu_e}] ~ \phi_{\nu_\mu}
+ P_{\nu_e \to \nu_\mu}~ \phi_{\nu_e}
\end{equation}
in the case  of $\nu_\mu \leftrightarrow  \nu_e$ oscillations,
and analogously for  antineutrinos.

Strictly speaking, the transition probability $P_{\nu_\mu \to \nu_x}$
depends
on the oscillation parameters $\Delta m^2$ and $\sin^2 2 \theta$,
neutrino energy $E_\nu$, nadir angle $\psi$, and also
 on the exact positions  of creation and absorption points
of neutrinos.
The neutrinos are produced in a layer of the atmosphere
with a thickness
of approximately $\sim 10$ km, and the  upward--going muons are created in  a
layer
of  thickness $\sim 0.1$ km below the detector.
We will  be  discussing a  range of $\Delta m^2/E$ such that
$\ell > 100$ km and therefore it is a good approximation
to  neglect  the dependence on the  exact position of creation and absorption
of
the neutrinos, and assume that all neutrinos
are  generated in an infinitely thin  spherical
layer of  atmosphere at the surface
of the earth and that the muons are also produced
in a thin  layer of material below the detector.
In this approximation the
$\nu_\mu \leftrightarrow \nu_\tau$ transition probability  takes the form:
\begin {equation}
P_{\nu_\mu \leftrightarrow \nu_{\tau}} \left ( { E\over \Delta m^2},
\cos \psi \right ) =
{1 \over 2} \sin^2 2 \theta \left [ 1 -
\cos \left ( {\Delta m^2 R_{\oplus} \cos \psi \over E } \right ) \right ].
\end{equation}
The probability of
$\nubar_\mu \leftrightarrow \nubar_\tau$ transition is equal to that in the
neutrino case.

The probabilities of the neutrino (and antineutrino)  transitions
$\nu_\mu \leftrightarrow \nu_e$ and
$\nu_\mu \leftrightarrow \nu_s$  have been calculated  by integrating
numerically the evolution equation (\ref{evolution}) using the
distributions of $N_e(r)$ and $N_n(r)$ obtained from the matter density
profile of the earth  taken from ref. \cite{Stacey} and assuming
$ N_n \simeq N_p = N_e $, which is quite a good approximation
in the interior of the earth.

\vspace {0.5 cm}
Some effects of neutrino oscillations on the
upward--going muon flux are described below.
(i) The  total  rate  is reduced with respect to expectations  based
 on the assumption of no oscillations.
To detect this effect one  needs to control
the absolute normalization of the  calculated flux.
(ii) The energy spectrum is  distorted, because low energy neutrinos
have a shorter oscillation length.
The shape of the neutrino spectrum, after the
crossing of the earth,  is reflected in the  energy spectrum
of the  upward--going muons. The ratio of stopping/passing muons
in an underground detector is  therefore a useful tool
in the  search for neutrino oscillations.
In this measurement  one needs only to control the shape (and not
 the normalization) of the calculated upward--going muon flux.
A large part of the   uncertainties in the calculations cancel,
and the measurement is thus dominated by statistical errors.
(iii) The nadir angle   distribution is also modified, because
longer path lengths correspond to more vertical directions.

\vspace {0.5 cm}
In our numerical  work we have  defined as ``stopping muons''
those  with energy
($1.25$ GeV $ \le E_\mu \le 2.5$ GeV), and
``passing muons'' those with $(E_\mu \ge 2.5$ GeV). Integrating
over the entire downward hemisphere
the muon flux given in eq. (\ref{upw-flux})
we obtain the integrated fluxes
\begin{equation}
\Phi^\pm_{s(p)} =  2\pi \int_0^1 d\cos \psi
{}~\int_{1.25 (2.5)}^{2.5(\infty)} dE_\mu ~\phi_{\mu^\pm}(E_\mu, \cos \psi).
\label{flux-muon}
\end{equation}
In general the oscillation probabilities of
neutrinos and antineutrinos are different, and therefore we   compute
separately the fluxes of positive and negative muons. However
the existing detectors are unable to
measure the charge of the  detected  particles, and so we then add together
the calculated $\mu^\pm$ fluxes.
The total flux of upward--going muons is
$\Phi = \Phi_s + \Phi_p = \Phi_s^+ + \Phi_s^- + \Phi_p^+ + \Phi_p^-$.

In the calculation of the
upward--going muon  fluxes in the absence of oscillations,   one needs to
specify  the neutrino fluxes, the  nucleon structure functions (needed in
the calculation of the cross section),
and the  muon range. The imperfect knowledge of the first and second
item introduce a significant systematic uncertainty.
Using the $(\nu_\mu + \nubar_\mu)$ flux calculated by Volkova \cite{Volkova80},
with the $\nu/\nubar$ ratio of Lipari \cite{Lipari},
the structure functions given by Diemoz {\it et al.} \cite{DFLM},  and the
muon range taken from the report of Lohmann {\it et al.} \cite{Lohmann},
the  calculated flux of upward--going muons  is:
\begin {equation}
{\Phi_s + \Phi_p \over 2 \pi} = 2.36 \cdot 10^{-13}
{}~{\rm (cm^2~s~sr)}^{-1}
\label{flutot_0}
\end{equation}
(0.70 of the flux is due to negative muons), and
\begin {equation}
{\Phi_s \over \Phi_p} = 0.185
\label{rstop_0}
\end{equation}
(0.67 of the  stopping muons are negative).
The median energy of neutrinos giving rise to stopping (passing) muons
is 5.5 (86) GeV.
A different choice of the input in equation (\ref{upw-flux}) would modify
these results,
in particular there is a large  sensitivity to the  neutrino fluxes.
Using the  neutrino fluxes of Butkevich {\it et al.} \cite{Butkevich},
results in a total upward--going muon flux
larger than  (\ref{flutot_0}) by 11\%. In this case the
stopping/passing ratio is  0.189, a difference of less
than 3\%.
Using the $(\nu_\mu + \nubar_\mu)$ fluxes of
 Mitsui \cite{Mitsui} would result in a  total flux
6\% (5.5\%) larger than (\ref{flutot_0}) and a ratio of stopping to passing
of 0.191 (0.190), with the $\nu/\nubar$ ratio of  Butkevich
(Lipari).
The calculations with the structure functions of Eichten {\it et al.}
\cite{EHLQ} result  in
a  rate of muons smaller by 1\% for set 1, and larger by 2.5\%
for set 2.

The calculated  upward--going muon flux  (\ref{flux-muon})
is implicitly  a function of the oscillation parameters.
Given the flavour of the neutrino  mixed with
the $\nu_\mu$,  each choice
of  a pair of parameters ($\Delta m^2$, $\sin^2 2 \theta$) will  result in
a different flux.
As an illustration, in fig. 2 we show the  calculated
$\Phi_s/\Phi_p$  ratio as a function of $\Delta m^2$ assuming
maximal mixing ($\sin^2 2 \theta = 1$).
It is  clear that  the matter effect is important and that the curves
corresponding to  $\nu_\tau$, $\nu_e$ and $\nu_s$   differ  from each other.
The qualitative behaviour of the curves  can be simply understood. For
$\Delta m^2  < 10^{-4} ~{\rm eV^2}$, the oscillation length even of the
softest neutrinos that  produce detectable muons  is  longer than the earth's
radius, and there is no visible effect. With increasing  $\Delta m^2$
the low energy neutrinos  begin to oscillate, the spectrum is distorted
and $\Phi_s/\Phi_p$ decreases.
For $\Delta m^2  > 1 ~{\rm eV^2}$, all relevant  neutrino energies are equally
affected  and, averaging over the very rapid
oscillations,  the shape of the spectrum is  again the same as in the
no-oscillation case.
In the case of \oscel ~oscillations we have to consider the presence of the
$\nu_e(\nubar_e)$ directly produced in meson decay. These neutrinos have a
much steeper  spectrum than the
$\nu_\mu(\nubar_\mu)$.
For large  $\Delta m^2$
averaging over rapid undetectable   oscillations we have:
$\phi_{\nu_\mu} \to (1- 0.5 \sin^2 2 \theta)\phi_{\nu_\mu }
+ 0.5 \sin^2 2 \theta~ \phi_{\nu_e}$. This yields a distortion of
the spectrum with a relative  excess of low energy  neutrinos.

\vspace {0.5 cm}
Measured values of the upward--going muon flux (averaged over  $2 \pi$ of
solid angle), in units $10^{-13}\,({\rm cm^2~s~sr})^{-1}$, are:
$(2.77 \pm 0.17)$ with a threshold of 1 GeV from
the Baksan group \cite{Baksan91},
$(2.26 \pm 0.17)$  with a threshold of 2 GeV from
the IMB experiment \cite{IMB90},
$(2.04 \pm 0.13)$  with a threshold of approximately
3 GeV from  the Kamiokande experiment \cite{KAM92b}.
The three measurements, after corrections for the
different thresholds, are in
agreement with each other, and very close to  our calculated value
(\ref{flutot_0}).
To obtain an  allowed region in the ($\Delta m^2, \sin^2 2 \theta$) plane
would require a detailed discussion of the systematic uncertainties
of the theoretical calculation.
The IMB  group \cite{IMB92b} derives a rather stringent limit
from their measurement of the total rate of upward--going muons.
A more critical view of the systematic uncertainties, however, would result
in much less stringent limit \cite {Frati}.
Rather arbitrarily,  we have chosen
as a criterion  to
define   the region of parameter space that can be excluded at 90\% c.l.
because of the measurement of the total rate the condition
\begin{equation}
\Phi \le 0.75~ \Phi_0\;,
\label{crit-A}
\end{equation}
$\Phi_0$ being the  average flux  (\ref{flutot_0})
calculated in the absence of oscillations.
The limiting curves calculated using this criterion
(curves of type $A$) for  different types of oscillations are shown
in fig. 3 for the \osctau ~and \oscster ~cases, and in
fig. 4 for the \oscel ~case.

\vspace {0.5 cm}
The IMB  collaboration  has measured \cite{IMB92b} a  stopping/passing ratio
$0.160 \pm 0.019$, in good agreement with their
detailed  Monte Carlo predictions.
Our calculated value of 0.185
agrees reasonably well with the IMB result,
considering the   approximations  we have
used\footnote{In the IMB detector
 the threshold energy for muon detection was  approximately
1.8 GeV  in the initial part of the data taking (IMB--1,2), and 1.0 GeV
for the  final part (IMB--3). The criterion for  stopping muons is that there
is
essentially no light for the last 5 meters of  the projected track length
inside  the detector. Considering the geometrical
dimension of the detector ($18 \times 17\times 22.5~m^3$) ,  this sets a
maximum energy for stopping muon of $\sim 2.5$ GeV.}.
In this case we have
used as a criterion (90\% c.l.) for oscillations
the condition:
\begin {equation}
{\Phi_s \over \Phi_p}  \le  0.8~ \left ( {\Phi_s \over \Phi_p} \right)_0
\label{crit-B}
\end{equation}
where again the  subscript
indicates the ratio of fluxes calculated  in the absence
of neutrino oscillations.
Curves  calculated with this criterion  (curves of type $B$) are shown
in fig. 3 for the \osctau ~and \oscster ~cases.
For the \oscel ~oscillations the condition (\ref{crit-B}) is  never satisfied.
In fig. 4 we show the  region of parameter  space that could be excluded with
the   requirement
$\Phi_s/\Phi_p  \le  0.9~ (\Phi_s/\Phi_p)_0$, more demanding from the point
of view of statistical accuracy.

\vspace {0.5 cm}
The  situation about experimentally allowed values of oscillation parameters
in the case of \osctau ~and \oscster ~oscillations
is summarized in fig. 3.
In this figure  three limiting curves are the results of experiments that do
{\it not} involve upward--going muons.
Curve (a)  is obtained
 from accelerator experiments  \cite{CDHS} ~and the allowed region
is below the curve.
Curve (b)  is obtained   from the
result of Fr\'ejus  \cite{Frejus}  on the
$e/\mu$ ratio of contained  events; this
result is consistent  with  the no--oscillation hypothesis,
and the allowed region is  to the left of the curve.
Curve (c)  is obtained   from the
observation of  an anomaly in the same $e/\mu$ ratio
by Kamiokande and IMB \cite{KAM92a,IMB92a};
 this is a `positive result' and the
allowed region is to the right of the curve.
These three limits apply  equally to
\osctau ~and \oscster ~oscillations which differ from each other only because
of the matter effects.
These effects are obviously irrelevant  to the
accelerator limit, and almost so also for (b) and (c) curves, although
approximately half of the contained events observed in
IMB and Kamiokande  are produced by   neutrinos that have
penetrated through the earth. In fact, these neutrinos have energies
$E_\nu \le 1.2$~GeV, with a rapidly   falling spectrum.
On curves (b) and (c) the absolute value of the squared mass
difference is always larger than $10^{-3}~{\rm eV}^2$. For matter effects to
be significant one needs
\begin{equation}
{\Delta m^2\over E} \aprle \sqrt{2} G_F N_A \rho
= 0.758 \cdot 10^{-4} \rho({\rm g\,cm^{-3})  ~ {eV^2 \over GeV}}
\end{equation}
The maximum density of the earth is  $\rho \simeq 12.5$ g\,cm$^{-3}$,
thereforefor contained events matter effects are  practically negligible
and their inclusion would only   slightly modify the shape of curve (c)
in the region of lowest $|\Delta m^2|$.

Matter effects are  on the contrary significant   for upward--going muons,
the reason being that
the neutrino energies involved  are
one to two orders of magnitude larger  than those for contained events.

The limits obtained  from the total flux of
upward--going muons using the criterion (\ref{crit-A})   are shown
in fig. 3
by the  curves $A_\tau$, $A_{s^+}$ and $A_{s^-}$,
the subscript indicating the flavour  of the neutrino mixed with
the $\nu_\tau$.
The case $\nu_s$   is described by two curves
because we need to consider the sign of
$\Delta m^2$.
The limits obtained  from the stopping/passing ratio
using the criterion (\ref{crit-B})   are shown
in the same figure by the curves $B_\tau$, $B_{s^+}$ and $B_{s^-}$.

Comparing the three type $A$ curves we can make the following remarks.
For maximal mixing  the limits on $|\Delta m^2|$  for
\oscster ~oscillations is
a factor 2.6 less stringent  than for \osctau.
If $\sin^2 2 \theta = 1$
the  matter can only suppress the oscillations, in fact
$\nu_\mu$'s with $|\Delta m^2|/E \aprle 2 \cdot
10^{-4}~{\rm eV^2/GeV}$ have their mixing  with  $\nu_s$
strongly suppressed (see fig. 1) and do not oscillate.
For maximal mixing  the sign of $\Delta m^2$ is irrelevant and therefore
the curves   $A_{s^+}$ and $A_{s^-}$  end at the same point.
With decreasing mixing
the two curves separate, the limit for
negative $\Delta m^2$ being stronger.
When $\Delta m^2 < 0$ ($\Delta m^2 > 0$) for oscillations of neutrinos
(antineutrinos)
the MSW resonant enhancement will occur.
Since $\nu_\mu$'s are more abundant (and have a larger cross section)
than $\nubar_\mu$'s, the same values of $\sin^2 2 \theta$ and  $|\Delta m^2|$
will result in a stronger suppression of the upward--going muon flux if
$\Delta m^2$ is negative and the MSW  resonance is relevant for neutrinos.
For $|\Delta m^2| \aprge 1~{\rm eV^2}$ the three curves $A_\tau$, $A_{s^+}$
and $A_{s^-}$ almost coincide; this is again a reflection of the fact that
for  large $|\Delta m^2|/E$  matter effects are not significant.

Very similar   considerations can be made about the curves
$B_\tau$, $B_{s^+}$ and $B_{s^-}$.
For maximal mixing,  the region of $|\Delta m^2|$ that can be excluded
in the case of \oscster ~oscillations
is moved  to  values  of $|\Delta m^2|$ larger by a factor 2.8  than in the
\osctau ~case.  For oscillations into sterile neutrinos,
when the mixing is smaller  than unity,  the  limit that applies
for $\Delta m^2 < 0$ is stronger than  for the other case. This is again
because when  the  MSW resonance occurs for neutrinos  (antineutrinos)
the sensitivity
to oscillations  is stronger (weaker). A detailed analysis must also take into
account the   fact that when the neutrinos resonate the oscillation length
passes trough a  maximum. This is the reason why
the limit  $B_{s^+}$ is stronger than the limit
$B_{s^-}$ in  a limited  region of parameter space.
For  maximal mixing the curves $B_{s^+}$ and $B_{s^-}$ end in the same points.

The most significant difference between the limits for
\osctau ~and \oscster ~oscillations   is in  the region of low
squared mass differences.
A region  of parameter  space
$|\Delta m^2| = (3 \div 7) \cdot 10^{-4} ~{\rm eV^2}$
and $\sin^2 2 \theta \ge 0.7$  is excluded for
$\nu_\tau$ but not for $\nu_s$ oscillations. This region is however
already excluded by curve (c), provided that the anomaly in the $e/\mu$ ratio
of contained events is due to neutrino oscillations.
As shown in fig. 3,
for both types of oscillations there is a region of parameter
space which is  compatible with all existing experimental measurements. The
range of allowed values  for the parameters is  in both cases roughly:
$0.4 \aprle \sin^2 2 \theta \aprle 0.7$ and
$2 \cdot 10^{-3}\;({\rm eV^2}) \aprle |\Delta m^2| \aprle 0.4\;({\rm eV^2})$.
The allowed region  for $\nu_s$ is somewhat smaller
but contains also a small   part (for large  positive
$\Delta m^2$) which is not allowed for $\nu_\tau$.
In future, with more data, the sensitivity
of the curves of type $B$ (which is determined by the statistical errors)
will improve, exploring   the low $|\Delta m^2|$ part of
the allowed region.
To improve the sensitivity of the curves of type $A$, one needs rather to
 reduce the theoretical systematic  uncertainties.
If these could be reduced to the level of 15\% (at 90\% c.l.) the entire
allowed region of parameter space would then be explored by
measurements of  upward--going muons and one would be able to either
confirm or disprove the neutrino oscillations as a solution to the
atmospheric neutrino puzzle.

\vspace {0.5 cm}
The  situation about the experimentally allowed values of
the  oscillation parameters
in the case of \oscel ~oscillations
is summarized in fig. 4. The results from the analysis of solar neutrino
experiments refer to lower values of $|\Delta m^2|$ and are not shown.
As in fig. 3, three limiting curves are the results of experiments that do
{\it not} involve upward--going muons.
Curve (a)  is the limit obtained from reactor experiments \cite{Gosgen},
curve (b) and (c) are  obtained from  measurements of the
$e/\mu$ ratio of  contained events
in the Fr\'ejus \cite{Frejus} and Kamiokande and IMB experiments
\cite{KAM92a,IMB92a}. As before, curve (b)
excludes the region to its  right, curve (c) the region to its left.

The limits obtained  from the measurement of the total upward--going
muon flux according to criterion (\ref{crit-A}) is shown by
curves $A_{e^+}$ and $A_{e^-}$, the  two curves referring to the sign of
$\Delta m^2$ as before. The region excluded by the curves of type $A$ is
well inside the region of parameter space already excluded by the G\"osgen
reactor experiment. Matter suppresses the oscillations more strongly
than in the \oscster ~case, since the difference in the
effective potentials is twice as large now, see eqs.
(\ref{v_e},\ref{v_mu},\ref{v_s}).
 We also  have to take into account the fact that in cosmic ray showers
a $\nu_e(\nubar_e)$ flux  is produced as well.
Both effects reduce the  sensitivity of the measurement to
\oscel ~oscillations.

As already mentioned, the condition (\ref{crit-B})  is never satisfied, and
 therefore no limit can be obtained from  the stopping/passing ratio.
In the case of
maximal mixing,  $\Phi_s/\Phi_p$  reaches  (for
$|\Delta m^2| = 6 \cdot 10^{-2} {\rm eV^2}$) a minimum value of 0.150,
only 19\% smaller than the  value calculated in the absence of oscillations.
Curves $B_{e^+}$ and
$B_{e^-}$  are calculated with the more demanding criterion
$\Phi_s/\Phi_p  \le  0.9~ (\Phi_s/\Phi_p)_0$. They are an indication of the
sensitivity that could be obtained  with a sample of data
approximately four times  the one collected by IMB.

As shown in fig. 4,
a small region of parameter space
$0.35 \aprle \sin^2 2 \theta \aprle 0.7$ and
$4\cdot 10^{-3}\;({\rm eV^2}) \aprle |\Delta m^2| \aprle 2 \cdot 10^{-2}
\;({\rm eV^2})$ is compatible with all existing experimental measurements.
Future experiments  on upward muon fluxes
will be able to explore only a part of this region.

\vspace {0.5 cm}
As we have seen, the oscillations of $\nu_\mu$ into
a sterile neutrino state $\nu_s$ can be
strongly affected by the matter of the earth.
Although the existence of a sterile
neutrino does not contradict any laboratory data, it can be in
conflict with cosmological considerations. Namely, for large enough
values of the mixing angle and $\Delta m^2$, $\nu_{\mu}\leftrightarrow
\nu_s$ oscillations can bring the sterile neutrinos into equilibrium with
matter before the nucleosynthesis epoch \cite{Dolgov} thereby affecting
the primordial ${}^4$He abundance in the universe.
The analysis performed in \cite{Chicago} sets the maximum allowed
number of ``light neutrinos'' $N_\nu$ to 3.4.
The limit derived in \cite{Enqvist}  not to have $N_\nu > 3.4$
for oscillations between sterile and muon neutrinos
 is $\Delta m^2 < 8\cdot10^{-6}\;{\rm eV}^2$ for maximal mixing.
With the values we obtained in our analysis this bound would be
violated, and we would have $N_\nu = 4$. The limit given in
\cite{Chicago} is not however universally accepted and is still
a point of debate \cite{Sarkar}, so that we believe that
$\nu_\mu \leftrightarrow \nu_s$ oscillations may still provide a viable
solution to the atmospheric neutrino problem.

As to the \oscel ~oscillations hypothesis for atmospheric neutrinos,
it should be noted that this assumption
does not contradict the idea of neutrino oscillations
being also responsible for the observed solar neutrino deficit. The
latter can be accounted for through the $\nu_e \leftrightarrow
\nu_{\tau}$ oscillations provided the neutrino masses obey the
conditions $m_{\nu_e}\approx m_{\nu_\tau} \ll m_{\nu_\mu}$ or
$m_{\nu_\mu} \ll m_{\nu_e}\approx m_{\nu_\tau}$. We would like
to stress that although many models (including those implementing
the see-saw mechanism \cite{seesaw}
of neutrino mass generation) predict the
neutrino mass hierarchy $m_{\nu_e}  \ll m_{\nu_\mu} \ll m_{\nu_\tau}$,
models with quite different hierarchies also exist.
Moreover, there are absolutely no experimental indications in favour
of the direct hierarchy.

In summary, we have shown that the complex of data induced by atmospheric
neutrinos (``contained'' and ``upward--going muons'') may still be
described in terms of neutrino oscillations. The matter effects are
important in the precise determination of the allowed parameter region.

\vspace{0.5cm}
P.L. is grateful to T.K. Gaisser and T. Stanev for many useful conversations
and for showing us an early version of ref. \cite{Frati}.
E.A. acknowledges fruitful discussions with Z. Berezhiani and A. Smirnov.

\newpage

\noindent {\bf Figure Captions}

\vspace {0.3 cm}
\begin{itemize}
\item [] Fig. 1. Plot of  $\sin^2 2 \theta_m$,
 the  mixing parameter in matter for $\nu_\mu \leftrightarrow \nu_e$
oscillations, as a function of
$\Delta m^2/E$, for $\sin^2 2 \theta = 1,0.5,0.05$, assuming a matter density
$\rho= 5 ~{\rm g\,cm^{-3}}$ and an electron fraction $Y_e = 0.5$.
Solid curves are for  neutrinos, dashed curves for antineutrinos.
If $\Delta m^2 <0$ the neutrino and antineutrino curves must
be interchanged.
 The same curves can also be interpreted as
the  mixing parameter in matter  for the $\nu_\mu \leftrightarrow \nu_s$
transition,  assuming a  matter density $\rho^{\prime} = 10~
{\rm g\,cm^{-3}}$
and interchanging the neutrino and antineutrino curves.

\item [] Fig. 2. Plot of the $\Phi_s/\Phi_p$ ratio as a function of
$|\Delta m^2|$, assuming  maximal  mixing in vacuum.
The different curves are for $\nu_\mu \leftrightarrow \nu_\tau$,
$\nu_\mu \leftrightarrow \nu_e$,
and $\nu_\mu \leftrightarrow \nu_s$ oscillations. We also show
the   curve calculated
for the $\nu_\mu \leftrightarrow \nu_e$ transitions
neglecting the  $\nu_e(\nubar_e)$ created directly in meson decays.

\item [] Fig. 3.
Limits for the oscillation parameters
$\Delta m^2$ and $\sin^2 2 \theta$ in the case of
$\nu_\mu \leftrightarrow \nu_\tau$ and
$\nu_\mu \leftrightarrow \nu_s$   mixing.
The curves $A_\tau$, $A_{s^+}$ and $A_{s^-}$ are
limits  obtained from the  measurements of the  total flux
of upward--going muons.
The curves $B_\tau$, $B_{s^+}$ and $B_{s^-}$ are
obtained from the measurements of the stopping/passing ratio.
Also plotted are the 90\% c.l. limits from accelerator experiments
(a),  and  the from the measurement of the
$e/\mu$ ratio of contained events in  the Fr\'ejus (b) and
Kamiokande experiments (c).

\item [] Fig. 4.
Limits for the oscillation parameters
$\Delta m^2$ and $\sin^2 2 \theta$ in the case of
$\nu_\mu \leftrightarrow \nu_e$   mixing.
The curves $A_{e^+}$ and $A_{e^-}$ are
obtained from the  measurements of the  total flux
of upward--going muons.
The curves $B_{e^+}$ and $B_{e^-}$ are in this case
estimates of the sensitivity of future measurements
of the stopping/passing ratio (see text).
Also plotted are the 90\% c.l.
limits from the G\"osgen reactor experiment (a), and from the
$e/\mu$ ratio of contained events in  Fr\'ejus (b),
and  Kamiokande (c).
\end{itemize}

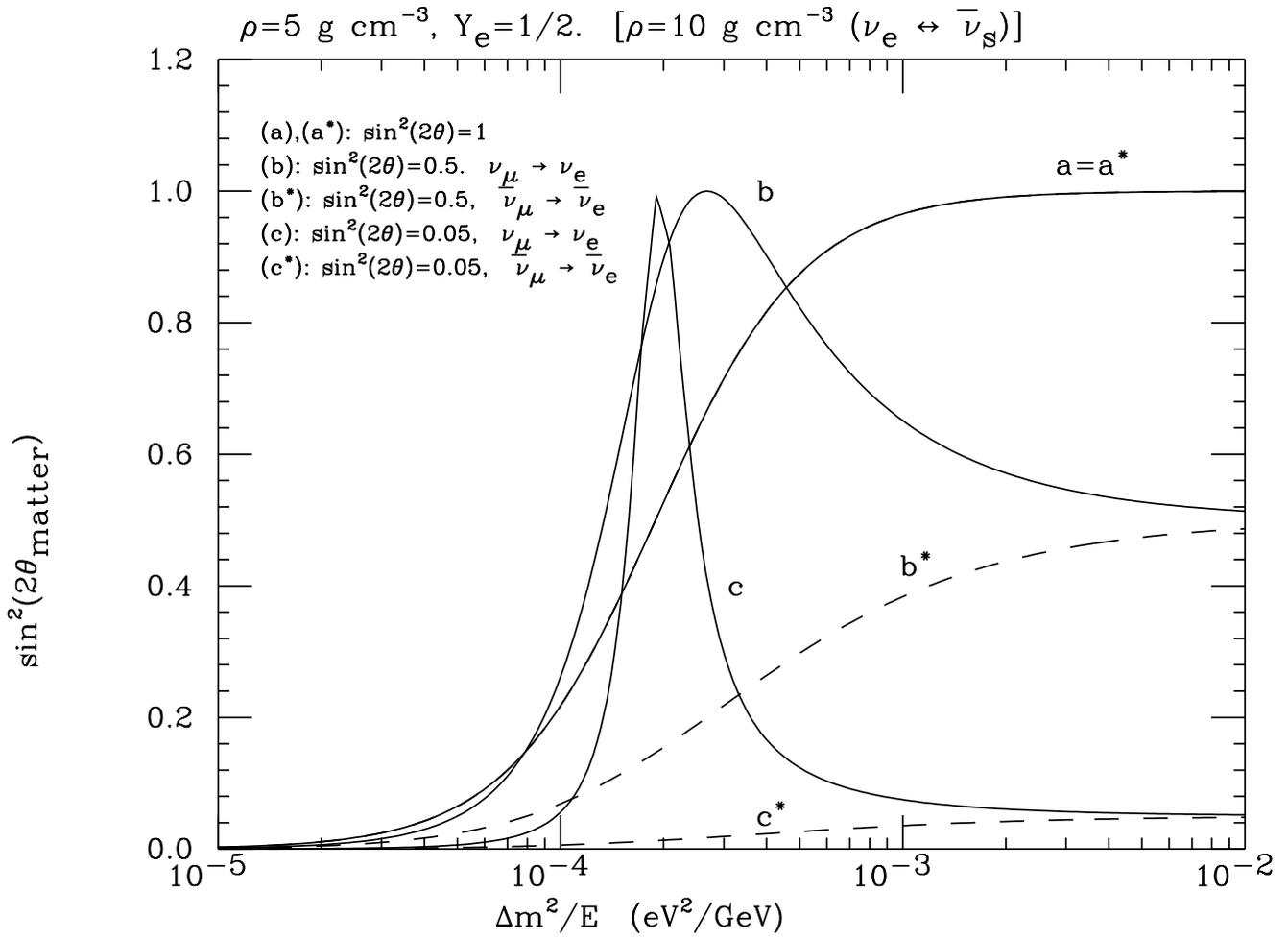
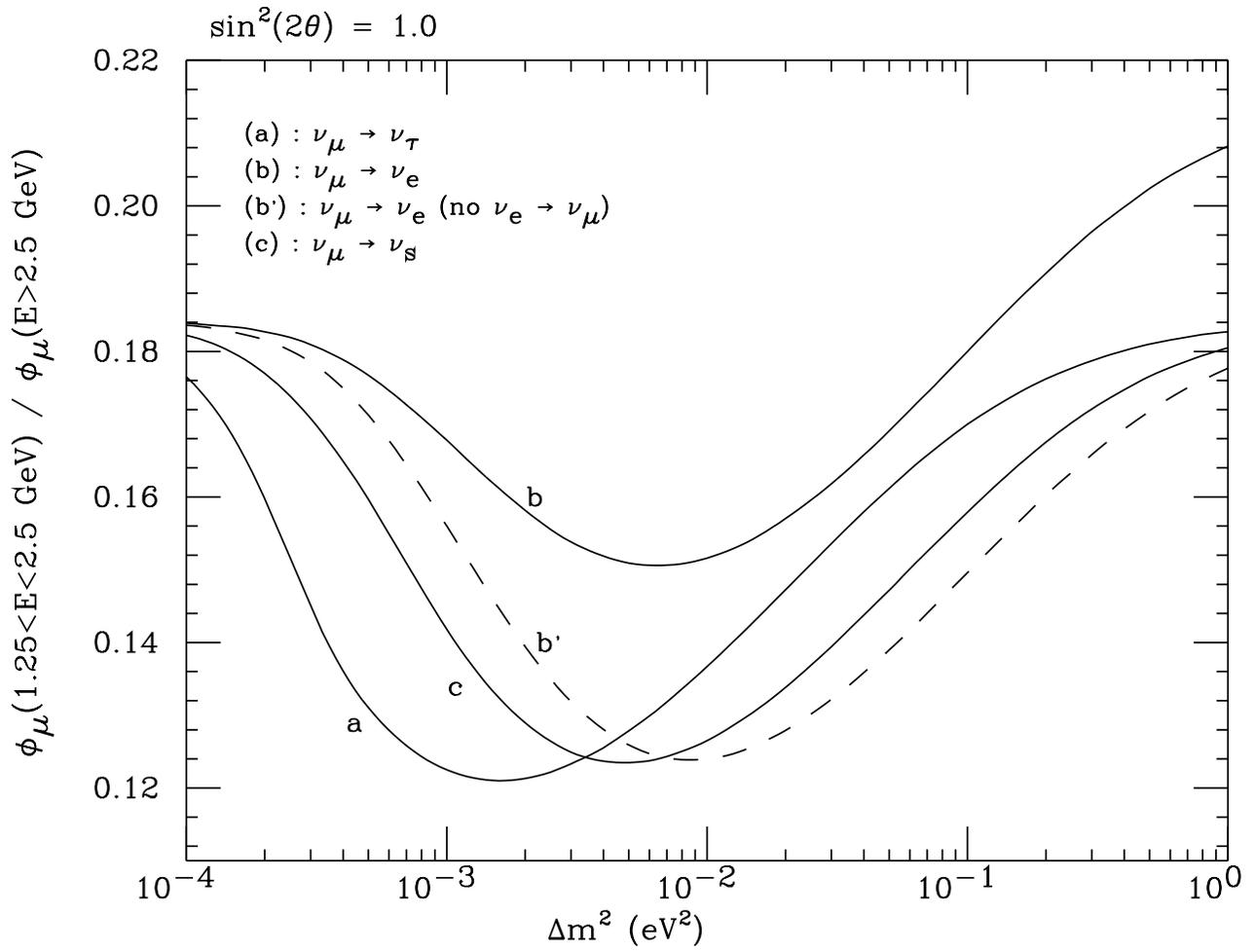
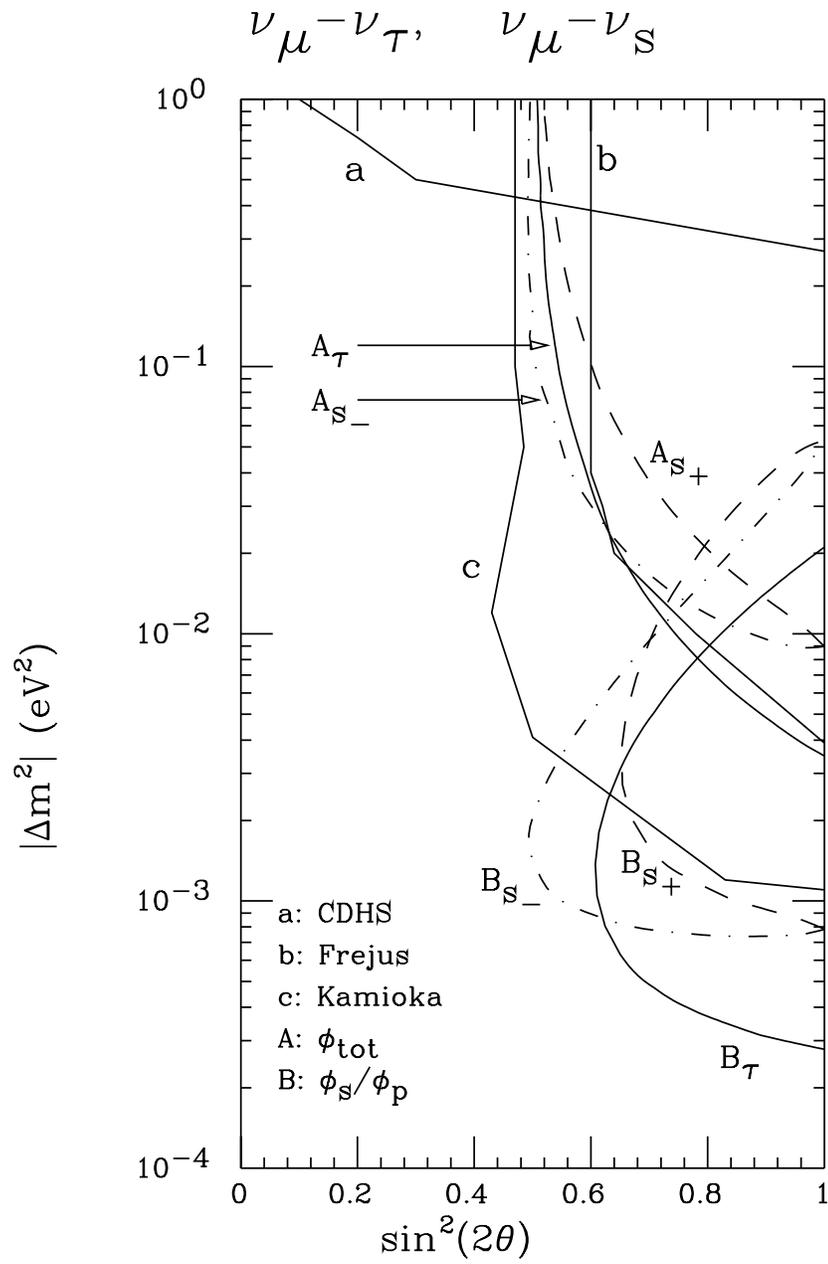
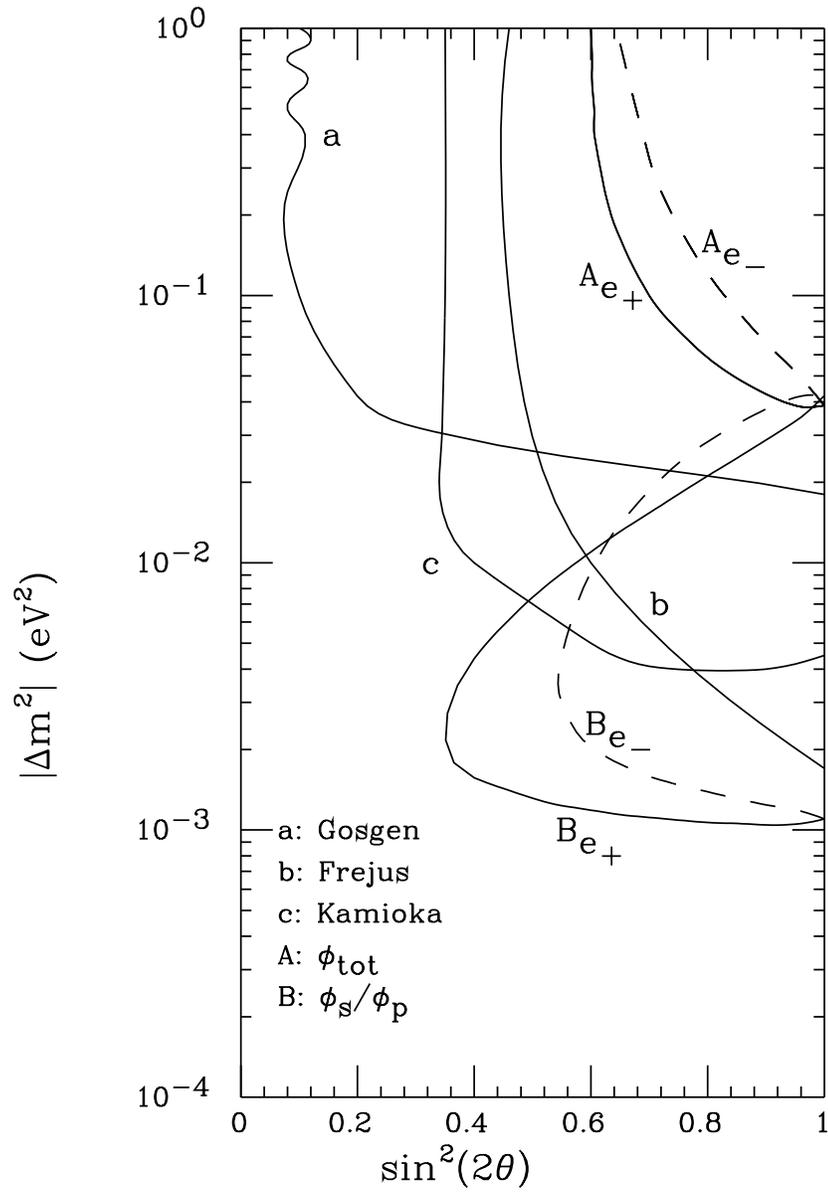
\end{document}